\documentclass[aps,prd,final,twocolumn,twoside,letterpaper]{revtex4}
\usepackage{amsmath}
\usepackage{amssymb}
\usepackage{amsthm}
\usepackage{enumerate}
\usepackage{array}
\usepackage{breqn}
\usepackage{tabularx}
\newcommand{\calP}{\mathcal{P}}

\newcommand{\calC}{\mathcal{C}}
\newcommand{\calS}{\mathcal{S}}
\newcommand{\calU}{\mathcal{U}}
\newcommand{\ket}[1]{\left|#1\right\rangle}
\newcommand{\bra}[1]{\left<#1\right|}
\newcommand{\braket}[2]{\left<#1|#2\right>}
\newcommand{\braopket}[3]{\left<#1\left|#2\right|#3\right>}
\newcommand{\cl}[1]{\textbf{Claim #1}}
\theoremstyle{remark}
\newtheorem{remark}[subsection]{Remark}
\theoremstyle{plain}
\newtheorem{theorem}[subsection]{Theorem}
\newtheorem{proposition}[subsection]{Proposition}
\theoremstyle{definition}
\newtheorem{definition}[subsection]{Definition}
\newtheorem{example}[subsection]{Example}

\begin{document}
\title{Extending the Graph Formalism to Higher-Order Gates}
\author{A. Khesin}
\author{K. Ren}
\affiliation{MIT}
%\date{\today}
\begin{abstract}
\fontsize{9pt}{9pt}\selectfont
We present an algorithm for efficiently simulating a quantum circuit in the graph formalism.
In the graph formalism, we represent states as a linear combination of graphs with Clifford operations on their vertices.
We show how a $\calC_3$ gate such as the Toffoli gate or $\frac\pi8$ gate acting on a stabilizer state splits it into two stabilizer states.
We also describe conditions for merging two stabilizer states into one.
We discuss applications of our algorithm to circuit identities and finding low stabilizer rank representations of magic states.
\end{abstract}
\fontsize{12pt}{12pt}\selectfont

\maketitle

\pagestyle{myheadings}
\markboth{Khesin and Ren}{Extending the Graph Formalism to Higher-Order Gates}
\thispagestyle{empty}
% \papersec{Introduction}
% Normal text example.
% Also I'll add the double spacing and other specifications of the phys rev tex details.
% \paperthm{NameOfTheorem}{Theorem example}
% Normal text and you can see how to cite a theorem by the result we saw in \thmNameOfTheorem.
% Also there are tons of other abbreviations like def, cnj, lem, and so on.
% Equations can be done
% \papereq{NameOfEquation}{like=this}{SourceHere, \cite{Dude}}
% and referenced like \eqNameOfEquation.

% There's a whole ton of other little quality of life stuff, but you won't even notice it.
% lmk if you need to add figures and tables,
% there's commands for that too

\section{Introduction}

Quantum logic gates play a crucial role in quantum informatics.
These gates allow a quantum circuit to behave predictably when presented with a certain quantum state by transforming this state into a desirable one.
While classical computations could all be carried out using a Turing machine, the physical reality of how classical circuits are implemented makes it useful to describe classical circuits in terms of gates.
Specifically, the NAND gate behaves similarly to a transistor and is therefore ``fundamental'', in some sense.
The NAND gate is in fact universal, meaning all Boolean circuits can be constructed solely out of NAND gates.

It is harder to describe the concept of universality in the quantum case because quantum operations are continuous and not discrete.
Notably, if we only limit ourselves to using a finite number of universal quantum gates, we will never be able to describe all possible unitary operations that can be done, as we could only ever make a countable number of circuits.
Universal quantum computation must be able to approximate unitary quantum operations with arbitrary precision.

As Aharonov showed in \cite{aha}, a small set of gates, such as the Clifford and Toffoli gates, allows for universal quantum computation.
It is thus desirable to not only be able to classically simulate how these gates act on quantum states, but also to do it quickly and in a small amount of space.
Notably, it is worth asking how much time and space will this simulation take as a function of the number of qubits being simulated as well as of the number of operations performed.

There is a classification of operations on quantum states called the Clifford hierarchy.
We call Pauli gates and Clifford gates $\mathcal{C}_1$ and $\mathcal{C}_2$, respectively.
The Toffoli gate is in a larger set of gates called $\mathcal{C}_3$.
We wish to be able to efficiently approximate universal circuits by simulating the action of $\mathcal{C}_3$ gates on quantum states.
We can accomplish this by considering only a particular set of quantum states called stabilizer states, which can be represented efficiently.
While Clifford gates take stabilizer states to stabilizer states, $\mathcal{C}_3$ gates take stabilizer states to linear combinations of two stabilizer states.

In this paper, we show how to simulate this action efficiently both in terms of time and space, as well as how to reduce the complexity of the representation by showing how to combine some linear combinations of two stabilizer states into one.
We accomplish this by providing explicit formulas and algorithms, and we state the conditions in which the above can be applied.
These results have important implications for questions regarding quantum computation and stabilizer rank.

In Sections II, III, and IV we introduce definitions of the Clifford hierarchy, stabilizer rank, and graph states, respectively.
In Section V we state our results and show examples of how to apply them in Section VI.
The proof of the main result is in Section VII and the runtime in computed in Section VIII.

\section{Clifford Hierarchy}

We now introduce a family of sets of matrices known as the \textit{Clifford hierarchy}.
These are key to the study of quantum circuits.

We begin by defining a fundamental set of unitary matrices, the Pauli group.
Let $n$ be the number of qubits we are simulating.

\begin{definition}\label{pauli}
Let $\calP$, the \textit{Pauli group}, be the group generated by the unitary matrices \{$X$, $Y$, $Z$\}, where $X=\left(\begin{smallmatrix}0 & 1\\ 1 & 0\end{smallmatrix}\right)$, $Y=\left(\begin{smallmatrix}0 & -i\\ i & 0\end{smallmatrix}\right)$, and $Z=\left(\begin{smallmatrix}1 & 0\\ 0 & -1\end{smallmatrix}\right)$.
We also let $I=\left(\begin{smallmatrix}1 & 0\\ 0 & 1\end{smallmatrix}\right)$. $\calP$ is on the first level of the Clifford hierarchy, so we also denote it $\calC_1$.
\end{definition}

The matrices $X$, $Y$, and $Z$ are called Pauli matrices, and have many nice properties.

\begin{remark}
The Pauli matrices obey $XY=iZ$, as well as similar expressions obtained by cyclically permuting $X$, $Y$, and $Z$.
Also note that the matrices $X$, $Y$, and $Z$ are often denoted $\sigma_x$, $\sigma_y$, and $\sigma_z$ or $\sigma_1$, $\sigma_2$, and $\sigma_3$, respectively.
\end{remark}

In general, if we wish to denote a gate $U$ applied to a specific qubit $a$ (or pair of qubits $a,b$), we will write $U_a$ (respectively, $U_{a,b}$) and it is implied that the tensor product is taken with identity matrices applied to all other qubits.
Furthermore, when writing tensor products of several matrices, we will usually omit the $\otimes$ sign for brevity.

Since unitary matrices represent quantum gates, it is useful to examine the conjugation of one matrix by another.
In other words, if a circuit contains the matrix $A$, we want to know whether the matrix $B$ commutes with it, and if it doesn't, what is the value of $C$ in the expression $BA=AC$.
Equivalently, we want to know what $B$ turns into if we ``pull it through'' $A$.
We find that $C=A^{-1}BA$.

It is now natural to examine the normalizer of $\calP$, the set of matrices $C$ through which we can pull a Pauli matrix and be sure to produce a Pauli matrix.

\begin{definition}\label{clifford}
Let $\calC_2$, the Clifford group, be the normalizer of $\calP$, given by $\calC_2=\{U\in \calU(2^n)\mid U^{-1}\calP^{\otimes n} U=\calP^{\otimes n}\}$.
\end{definition}

As Gottesman shows in \cite{got}, $\calC_2$ is generated by the Hadamard gate, $H=\frac1{\sqrt2}\left(\begin{smallmatrix}1&1\\1&-1\end{smallmatrix}\right)$, the phase gate $S=\left(\begin{smallmatrix}1&0\\0&i\end{smallmatrix}\right)$, and any controlled Pauli gate, such as the controlled-$X$ gate, $CX=\left(\begin{smallmatrix}1&0&0&0\\0&1&0&0\\0&0&0&1\\0&0&1&0\end{smallmatrix}\right)$, which applies the $X$ gate to the target qubit if the control qubit is in the $\ket1$ state. When using subscripts to indicate the qubits to which the gate is applied, we write the control qubit first. The above matrix shows $CX_{1,2}$. 

\begin{example}
The $\textit{SWAP}$ gate is a $\calC_2$ gate formed by $CX_{1,2}\cdot CX_{2,1}\cdot CX_{1,2}$, which swaps the values of qubits 1 and 2 and is written as $\left(\begin{smallmatrix}1&0&0&0\\0&0&1&0\\0&1&0&0\\0&0&0&1
\end{smallmatrix}\right)$.
\end{example}

\begin{definition}\label{localClifford}
We define a \textit{local Clifford operation} as any of the 24 matrices in $\langle H,S\rangle$, the set of matrices generated by multiplying a sequences of $H$ and $S$ matrices, which are all of the possible Clifford operations that can be applied to a single qubit.
\end{definition}

We can further extend this idea and ask ourselves which operators map Pauli gates to the Clifford group under conjugation.
This leads us to the following definition.

\begin{definition}\label{hierarchy}
For $k\in\mathbb{N}$, let $\calC_k$, the \textit{operators on level $k$ of the Clifford hierarchy}, be defined recursively as $\calC_k=\{U\in \calU(2^n)\mid U^{-1}\calP U\subseteq \calC_{k-1}\}$, with the base case being $\calP=\calC_1$. Note that the Clifford group is indeed $\calC_2$ under this definition. Furthermore, we note that for $k>2$, these operators do not form a group.
\end{definition}

%To achieve universal quantum computation, we need to be able to approximate any unitary matrix with arbitrary precision.
As we mentioned earlier, we can achieve universal quantum computation by just using Clifford gates and the Toffoli gate, which is in $\calC_3$.

\begin{remark}\label{cOp}
To apply a controlled operation $U$, the matrix used is $\left(\begin{smallmatrix}I&0\\0&U\end{smallmatrix}\right)$. The operation $U$ is controlled by qubit 1 and applied to remaining qubits. This operation is denoted $CU$, which can be nested if $U$ is already a controlled operation. Note that applying a control operation to a gate raises its level in the Clifford hierarchy. For the rest of the paper, we use the convention that if a controlled operation $CU_{a,b}$ is applied to qubits $a$ and $b$, then when $b=a$ this operation is equal to $U_a$.
\end{remark}

The gates in $\calC_3$ that we study are $T=\left(\begin{smallmatrix}1&0\\0&e^{\frac{i\pi}4}\end{smallmatrix}\right)$, $CH$, $CS$, $CCX$, $CCZ$, $\textit{CSWAP}$. These are the $\frac\pi8$ gate, the controlled-$H$ gate, the controlled-$S$ gate, the Toffoli or controlled-controlled-$X$ gate, the controlled-controlled-$Z$ gate, and the controlled-SWAP gate, respectively.
For ease of notation we denote the set of these six $\calC_3$ operators as $\calS$.

\section{Stabilizer Rank}

When doing quantum simulations, it is convenient to start the simulation with a particular state, such as $\ket{0}^{\otimes n}$, and then proceed from there.
In fact there is a specific finite subset of all $n$-qubit states which is convenient to work with.
For this we introduce the stabilizer formalism.

\begin{definition}
We say that a state $\ket\psi$ is a \textit{stabilizer state} if there exists a set of $n$ independent operators $U\in\calP^{\otimes n}$ such that $U\ket\psi=\ket\psi$.
We say that the operators $U$ stabilize $\ket\psi$.
\end{definition}

%Stabilizer states are defined in terms of $\calP=\calC_1$ and have many useful properties relating to the Clifford hierarchy.

\begin{remark}
If $\ket\psi$ is a stabilizer state and $U\in\calC_2$, then $U\ket\psi$ will also be a stabilizer state. Furthermore, any stabilizer state can be expressed as $U\ket{0}^{\otimes n}$ for some $U\in\calC_2$.
\end{remark}

One of the things that make stabilizer states useful is that they can be represented efficiently.
An $n$-qubit state requires $2^n$ components to represent as a state vector, but the $n\times n$ stabilizer matrix of Pauli matrices that identify the stabilizer state only requires $O(n^2)$ space. This exponential improvement is particularly important if the number of qubits is large.

Bearing our goal of universal quantum computation in mind, the natural question that arises now that we have defined stabilizer states is how do $\calC_3$ gates act on these states.
Immediately, we see that if $\ket\psi$ is a stabilizer state and $U\in\calC_3$, then $U\ket\psi$ will not always be a stabilizer state, as that is only true for gates in $\calC_2$.
In fact, representing the action of several $\calC_3$ gates on $\ket\psi$ is of great interest as this lets us perform universal quantum computation, magic state distillation, and much more.
For now, we need to introduce new terminology to talk about the outputs of $\calC_3$ gates on stabilizer states.

\begin{definition}\label{stabRank}
As defined by \cite{bss}, let the \textit{stabilizer rank} $\chi(\ket\psi)$ of an $n$-qubit state $\ket\psi$ be the smallest number $k$ of stabilizer states $\ket{\phi_i}$ needed to write $\ket\psi=\sum\limits_{i=1}^kc_i\ket{\phi_i}$.
Note that $\chi(\ket\psi)\geq1$ with equality iff $\ket\psi$ is a stabilizer state. Also we observe that $\chi(\ket\psi)\leq2^n$, since the standard basis vectors are all stabilizer states.
\end{definition}

We will later show that if $\ket\psi$ is a stabilizer state and $U \in \calS$ is one of the $\calC_3$ gates, then we can always write $U\ket\psi$ as a linear combination of no more than 2 stabilizer states.
In other words, $\chi(U\ket\psi)\leq2$.

This is a very important point, as this means that we can represent the action of a $\calC_3$ gate in a circuit as a linear combination of two stabilizer states.
Since matrix multiplication is additive, a second $\calC_3$ gate would act on both parts independently.
This means that we have reduced the problem of finding out how a series of $\calC_3$ gates acts on a stabilizer state, to learning the behaviour of just one.
The cost we pay for this is that our memory system must be able to store a linear combination of stabilizer states.

Presently, this means that applying $m$ $\calC_3$ gates will result in a linear combination of $2^m$ stabilizer states, which we would like to avoid.
What we focus on is detecting when a linear combination of two stabilizer states can be merged into one.
By applying such a merging operation after every application of a $\calC_3$ gate, we can massively reduce the exponential growth in the space required to represent our state.
However, with our current representation of stabilizer states, this merging operation would require us to identify when a linear combination of two stabilizer matrices can be merged and produce a stabilizer matrix of the sum.
We are not currently aware of a method that would allow us to do this, which is why we have to turn to a different representation of stabilizer states.

\section{Graph States}

A lot of work in quantum cryptography has been done using codes based on graphs.
Similar techniques allow us to adapt graphs for our purposes.

\begin{definition}\label{graph}
A \textit{graph} $G=(V,E)$ consists of a set of vertices $V$ and a set of edges $E\subseteq V\times V$ connecting those vertices.
In this paper we always take $V$ to be the set of $n$ qubits numbered from $1$ to $n$.
By $\text{nbhd}(a)$ we denote the neighbourhood of $a$, the set of all vertices $b$ such that $(a,b)\in E$.
\end{definition}

We can associate a specific stabilizer state with each graph.

\begin{definition}\label{graphState}
Let the \textit{graph state} of a graph $G$, $\ket G$, be given by the state stabilized by the operators $g_i = X_i\prod\limits_{j\in\text{nbhd}(i)}Z_j$, $\forall i\in V$.
Note that $\ket G$ is a stabilizer state.
\end{definition}

However, not all stabilizer states can be expressed as graph states, which means we need to supplement our graphs with additional information.
As Van de Nest et.~al.~show in \cite{vdn}, any stabilizer state can be expressed as a graph state with local Clifford operations performed on the vertices (see Definition \ref{localClifford}).
This means that any stabilizer state can be denoted by $U\ket G$ where $U\in\langle H,S\rangle$ and $G$ is a graph.
Storing such a graph takes $O(1)$ additional space per qubit, as the number of local Clifford operations is fixed.
Storing the entire graph as an adjacency list takes $O(nd)$ space where $d$ is the average degree of the graph.

Now what we have to do is figure out when and how we can merge a linear combination of two graphs into one.

\section{Statement of results}

We now present our main result. Let $\calS$ be $\{ T, CS, CH, CCZ, CCX, \textit{CSWAP} \}$, a collection of $\calC_3$ gates.

\begin{theorem}\label{main}
Given an operator $C \in \calS$ and a state $U \ket{G}$ with largest degree $d$, we can find two states $U_1 \ket{G_1}$ and $U_2 \ket{G_2}$ such that $C U \ket{G} = U_1 \ket{G_1} + U_2 \ket{G_2}$ in runtime $O(nd)$, where $d$ is the average degree of $G$.
\end{theorem}
% To show this, we introduce the notion of separable projector.
% \begin{definition}
% A generalized 
% \end{definition}
% \begin{proposition}\label{sep}
% Every operator in $\calS$ is the sum of two projector products.
% \end{proposition}
% We also need to show that a separable projector acting on a stabilizer state gives a stabilizer state. Thus, we will need the following merge algorithm.

% \begin{proposition}\label{proj}
% The action of a one-qubit projector on a stabilizer state is a stabilizer state.
% \end{proposition}
We will prove Theorem \ref{main} by means of the following algorithm, generalizing Prop. 1 of \cite{heb}.
\begin{proposition}\label{merge}
Let $A=\text{nbhd}(1)\cup\{1\}$, and let $B$ be a set including $1$. Let $k$ be an integer. Then
\begin{multline}\label{initassume}
    \frac{1}{\sqrt{2}} \left(I + i^k \prod_{j \in B} Z_j \right) \ket{G} \\= H_1 Z_1\hspace{-1.5ex}\prod_{x \in A, y \in A}\hspace{-1.5ex}CS_{x,y}^k\hspace{-1.5ex}\prod_{x\in A, y \in B}\hspace{-1.5ex}CZ_{x,y} \ket{G}.
\end{multline}
(Note the convention from Remark \ref{cOp}.)
\end{proposition}

Note that the expression in Proposition \ref{merge} may contain a $CS_{x,y}$ operator, which is a $\calC_3$ gate that we are not allowed to apply to a graph.
We note that the product of such terms has terms $x\in A$ and $y\in A$.
Thus, for every term $x=a\neq y=b$, we will apply both $CS_{a,b}$ and $CS_{b,a}$ which multiply to $CZ_{a,b}$, the toggling of an edge.
Meanwhile, if $x=a=y$, then by our convention $CS_{a,a}=S_a$, which we are also allowed to apply.

% We will then apply Theorem \ref{main} to analyze several circuits of $\calC_3$ operators.

% \section{Algorithm}
% We will describe the algorithm to prove Theorem \ref{main}. We first restate Proposition \ref{sep}.
% \begin{proposition}
% Every operator in $\calS$ is the sum of two separable projectors.
% \end{proposition}

% \textit{Proof.} By inspection.
We will describe the algorithm to prove Theorem \ref{main}. First, Table \ref{decomp} allows us to decompose gates in $\calS$ into the sum of two products of projectors and $\calC_2$ gates:
\begin{table}[h]
\begin{center}
\fontsize{11pt}{16pt}\selectfont
\noindent\begin{tabular}{|c|c|}\hline
    \fontsize{9pt}{16pt}\selectfont$\mathbf{\calC_3}$ \textbf{Gate} & \textbf{Decomposition}\\\hline
    T   & $\frac12(I+Z)\phantom{I}+\frac12e^{\frac{i \pi}{4}}(I-Z)\phantom{S}$\\\hline
    CS  & $\frac12(I+Z)I+\phantom{e^{\frac{i \pi}{4}}}\frac12(I-Z)S$ \\\hline
    CH  & $\phantom{H}\frac12(I+Z)I+\phantom{e^{\frac{i \pi}{4}}}\frac12(I-Z)H\phantom{S}$ \\\hline
    CCZ & $III-\frac14(I-Z)(I-Z)(I-Z)$ \\\hline
    CCX & $III-\frac14(I-Z)(I-Z)(I-X)$ \\\hline
    \fontsize{9pt}{16pt}\selectfont \textit{CSWAP} & \fontsize{9pt}{16pt}\selectfont$III - \frac14(I-Z)\otimes (II-ZZ)\cdot(II-XX)$ \\\hline
\end{tabular}
\end{center}
\caption{A decomposition of the $\calC_3$ gates we examined into a sum of two products of projectors and $\calC_2$ gates. Note that all tensor products have been removed for brevity except for that in the final entry, to highlight the distinction between matrix product and tensor product.}
\label{decomp}
\end{table}

Here, a projector is a term of the form $I + P$, where $P \in \calP^n$.
The projectors have the nice property that conjugation by a local Clifford operator gives another projector. For example,
\begin{equation}
    H (I + Z) H = I + X.
\end{equation}
Thus, for any local Clifford gate $U$ and projector $P$, we have
\begin{equation}
    P U \ket{\psi} = U P' \ket{\psi},
\end{equation}
where $P'$ is another projector. With the local Clifford operations out of the way, we can now apply our merging algorithm to $P' \ket{\psi}$. Splitting $P' = P_1 P_2 \dots P_k$, a product of projectors, we will apply $P_k, P_{k-1}, \dots, P_1$ in order to $\ket\psi$.
Now we need to outline how to apply a single projector $P$ to $\ket\psi$.

If $P$ is a product of $Z$ gates, then we can directly apply the merging algorithm. Otherwise, we convert $Y = iXZ$ to replace all $Y$'s, and we can use the identity
\begin{equation}\label{xtoz}
    X_a\ket{G}=\left(\prod\limits_{b\in\text{nbhd}(a)}Z_b\right)\ket{G}
\end{equation}
to convert all $X$ gates to $Z$ gates on other qubits. This identity is valid because $X_a \prod\limits_{b\in\text{nbhd}(a)}Z_b$ is a stabilizer for the graph state $\ket{G}$. This operation might cause $Z$ gates to appear on other qubits that still had $X$ gates that needed to be applied to the graph. In this case, we use the anti-commutativity of the Pauli matrices to fix the order. Thus, we can make sure there are only $Z$ gates left, and we can apply the merging algorithm. The application of $P_1$ is thus another stabilizer state, and then we can apply $P_2$, $P_3$, and so on. This proves Theorem \ref{main}.

\section{Examples}
One important concern is that our algorithm behaves well with circuit identities. For example, applying two $CCX$ gates gives the identity, but each $CCX$ would turn one stabilizer state into two, so that we end up with four stabilizer states. A robust algorithm would find a way to merge or otherwise annihilate these four stabilizer states such that only one remains. This is the content of the next two examples, which deal with the identities $CS \cdot CS = CZ$ and $CCX \cdot CCX = III$. In both cases, we should expect the output to be a single stabilizer state. The third example is related, in which we show that the stabilizer rank of the 2-qubit magic state $TT\ket{++}$ is 2 and not 4.
\subsection{CS}
Start with a stabilizer state $\ket\psi = U\ket{G}$. Then
\begin{equation}
    \fontsize{10pt}{10pt}\selectfont CS \ket\psi = \frac{1}{2} \left( (I+Z)I + (I-Z)S \right) U\ket{G}.
\end{equation}
If we apply $CS$ again then we get
\begin{equation}
    \fontsize{10pt}{10pt}\selectfont CZ \ket\psi = \frac{1}{2} \left( (I+Z)I + (I-Z)S \right)^2 U\ket{G}.
\end{equation}
Now, we use the identity $(I+Z)(I-Z) = 0$ and $(I+Z)^2 = 2(I+Z)$ to simplify this to
\begin{equation}
    \fontsize{11pt}{11pt}\selectfont CZ \ket\psi = \left( (I+Z)I + (I-Z)Z \right) U\ket{G}
\end{equation}
We should have four stabilizer states, but two of them (corresponding to the cross terms) collapsed to zero. We can now perform merges on $(I+Z) U\ket{G}$ and $(I-Z) U\ket{G}$. According to the algorithm, we should first make them into
\begin{equation}
    U \left(I\pm i^k\prod_{q \in B} Z_q \right) \ket{G}.
\end{equation}
Without loss of generality, let $1 \in B$. Then the output of the algorithm will be two stabilizer states $U_1 \ket{G'}$ and $U_2 \ket{G'}$, where
\begin{equation}
    U_1 = U_2\prod_{a \in A} Z_a.
\end{equation}
The extra $Z_a$ terms come from the powers of $CS$ in Proposition \ref{merge}.
Thus, we can merge $U_1 \ket{G'}$ and $U_2 \ket{G'}$ into a single final state. As expected, two $CS$ gates produce a single stabilizer state.

\subsection{CCX}
Start with a stabilizer state $\ket\psi = U\ket{G}$. Then
\begin{multline}
    CCX \ket\psi =\\\left( III - \frac{1}{4} (I-Z)(I-Z)(I-X) \right) U\ket{G}.
\end{multline}
Apply $CCX$ again to get
\begin{equation}
    \fontsize{10pt}{10pt}\selectfont\ket\psi = \left( III - \frac{1}{4} (I-Z)(I-Z)(I-X) \right)^2 U\ket{G}.
\end{equation}
Let $U = (I-Z)(I-Z)(I-X)$. Then $U^2 = 8U$. Hence
\begin{equation}
    \left(I - \frac{1}{4} U \right)^2 = I - \frac{1}{2} U + \frac{1}{16} U^2 = I.
\end{equation}
We thus get the desired result.

To sum up, after the first $CCX$ we get two stabilizer states. After the second, we get four states, of which three combine to zero.

\subsection{Magic states}
We can simulate the application of $T$ gates. Consider $T_1 T_2 \ket{++}$, where $\ket{++}$ is the graph state of the empty graph on 2 vertices. As expected, one application of the algorithm leads to two stabilizer states:
\begin{equation}
    \fontsize{11pt}{11pt}\selectfont T_1 \ket{++} = \frac{1}{\sqrt{2}} H_1 \ket{++} + \frac{1}{\sqrt{2}} e^{\frac{i \pi}{4}} H_1 Z_1 \ket{++}.
\end{equation}
Applying the algorithm again, we get four stabilizer states:
\begin{multline}
    \fontsize{10pt}{10pt}\selectfont T_2 T_1 \ket{++}= \frac{1}{2} H_1 H_2\ket{++}+\\
    \fontsize{10pt}{10pt}\selectfont   \frac{1}{2} e^{\frac{i \pi}{4}} H_1 H_2 Z_2 \ket{++}+\frac{1}{2} e^{\frac{i \pi}{4}} H_1 H_2 Z_1 \ket{++} + \\
        \fontsize{10pt}{10pt}\selectfont  \frac{i}{2} H_1 H_2 Z_1 Z_2 \ket{++}.
\end{multline}
We can apply our merge on the first and fourth states, and the second and third states, to get
\begin{equation}
    \fontsize{11pt}{11pt}\selectfont T_2 T_1 \ket{++} = \frac{1}{\sqrt{2}} S_1 H_2 \ket{G} + \frac{1}{\sqrt{2}} e^{\frac{\pi i}{4}} H_1 Z_1 \ket{G},
\end{equation}
where $G$ is connected on two vertices, so $\ket G$ is $CZ\ket{++}$. We obtain a decomposition into two stabilizer states, which agrees with the stabilizer rank of $2$ found in \cite{bss}.

\section{Proof of Proposition \ref{merge}}
\subsection{An initial simplification}
We now show that if we prove Proposition \ref{merge} for the case of $k\in\{0,1\}$, then the $k + 2$ case follows. Let $A' = A\backslash\{ 1 \}$ and $B' = B\backslash\{ 1 \}$.
We assume that for $k\in\{0,1\}$, the equation in Proposition \ref{merge} holds.
Then, we use the following facts:
\begin{enumerate}[A.]
    \item $H_1$ commutes with $Z_p$ for $p \neq 1$, and $H_1 Z_1 H_1 = X_1$;
    \item Equation \ref{initassume}, the merge for the case $k$,
    \item Equation \ref{xtoz}, that allows us to convert $X_a$ to $Z_b$'s on $\text{nbhd}(a)$.
\end{enumerate}
\begin{multline}
    H_1 Z_1 \hspace{-1.5ex}\prod_{x \in A, y \in A} \hspace{-1.5ex}CS_{x,y}^{k+2} \prod_{p\in A, q \in B} \hspace{-2.5ex}CZ_{p,q} \ket{G} \\
        \overset{A}{=}X_1 \hspace{-0.5ex}\prod_{p \in A'}\hspace{-0.5ex} Z_p^k H_1 Z_1\hspace{-2ex} \prod_{p \in A, q \in A} \hspace{-2ex}CS_{p,q}^k\hspace{-2ex} \prod_{p\in A, q \in B} \hspace{-2ex}CZ_{p,q} \ket{G} \\
        \overset{B}{=}X_1 \prod_{p \in A'} Z_p^k \frac{1}{\sqrt{2}} \left(I + i^k \prod_{q \in B} Z_q \right) \ket{G} \\
        =\prod_{p \in A'} Z_p^k \frac{1}{\sqrt{2}} \left(I - i^k \prod_{q \in B} Z_q \right) X_1 \ket{G} \\
        \overset{C}{=}\prod_{p \in A'} Z_p^k \frac{1}{\sqrt{2}} \left(I - i^k \prod_{q \in B} Z_q \right) \prod_{p \in A'} Z_p^k \ket{G} \\
        =\frac{1}{\sqrt{2}} \left(I + i^{k+2} \prod_{q \in B} Z_q \right) \ket{G}
\end{multline}
Thus, we need to show this where $k\in\{0,1\}$.

\subsection{Outline of Proof}
To prove Proposition \ref{merge}, we need to verify that the input state and the output state have the same stabilizer group and global phase.

We remark that a similar result is given as Proposition 1 in \cite{heb}. However, they only prove the first of these claims. In the following, let $\zeta$ be the CZ's and $\sigma$ be the CS's in Equation \ref{initassume}.

\subsection{Same stabilizer group}
We first consider $k = 0$. Let $\ket{G}$ be stabilized by $g_1, g_2, \dots, g_n$. Let $\ket{\psi} = \left(I + \prod\limits_{j \in B} Z_j \right) \ket{G}$. Then applying the local Clifford operators in the RHS of Equation \ref{initassume} transforms the stabilizers by conjugation into $g_1', g_2', \dots, g_n'$. Thus, we know that $g_1'$ through $g_n'$ stabilize the RHS of Equation \ref{initassume}. We will later prove they also stabilize $\ket\psi$ by first proving three claims. Define $A' = A\backslash\{1 \}$ and $B' = B\backslash\{ 1\}$.
\begin{enumerate}
    \item $g_1' = \prod\limits_{j \in B} Z_j$.
    
    \item If $m \notin A$, then $g_m'$ stabilizes $\ket{G}$.
    
    \item If $m \in A'$, then $g_1' g_m'$ stabilizes $\ket{G}$.
\end{enumerate}

Before we prove the claims, we should mention a fact. The action of $CZ_{a,b}$ on a stabilizer $g_k$ is trivial (i.e. doesn't affect the stabilizer) if $k \neq a$ or $k \neq b$, and it is multiplication by $Z_b$ if $k = a$, and multiplication by $Z_a$ if $k = b$.

To prove \cl{1}, we compute the action of $H_1 Z_1 \zeta$ on $g_1$. This is represented in the following equation.
\begin{multline}
    g_1' = (-Z_1 X_1)  \cdot \prod\limits_{p \in A'} Z_p \prod\limits_{q \in B'} Z_q \\\cdot (-1) \cdot \left( X_1 \prod\limits_{p \in A'} Z_p \right) 
        = \prod_{q \in B} Z_q.
\end{multline}
All terms except the first and last are the action of $\zeta$:
\begin{itemize}
    \item the $CZ_{1,q}$ and $CZ_{p,1}$ multiply $g_1$ by $Z_q$ or $Z_p$, respectively,
    \item the $CZ_{p,q}$ for $p, q \neq 1$ do nothing,
    \item the $CZ_{1,1}$ is conjugation by $Z_1$, which turns $X_1$ into $-X_1$, hence the $-1$.
\end{itemize}
The first term represents the action of $H_1 Z_1$, which turns the $X_1$ into a $-Z_1$ (hence the multiplication by $-Z_1 X_1$).

\cl{2} has two cases: $m \in B$ and $m \notin B$.

For $m \notin B$, we have $g_m' = g_m$, because neither $\zeta$ nor $H_1 Z_1$ affects $g_m$.

For $m \in B$, the action of $\zeta$ is just $\prod_{p \in A} Z_p$, since $p \notin A$. Note that the first entry in $g_m$ was $I_1$, which is now converted to $Z_1$. The action of $H_1 Z_1$ converts the $Z_1$ to $X_1$. Thus $g_m' = (X_1 Z_1) \prod_{p \in A} Z_p g_m = g_1 g_m$, which stabilizes $\ket{G}$.

\cl{3} has two cases: $m \in B$ and $m \notin B$.

For $m \notin B$, the action of $\zeta$ gives $\prod_{q \in B} Z_q g_m$.
Note that the first qubit is $I_1$ because the edge $(1,m)$ is toggled off. Hence, $H_1 Z_1$ acts trivially.
Thus, $g_m' = \prod_{q \in B} Z_q g_m$, and so $g_1' g_m' = g_m$.

For $m \in B$, the action of $\zeta$ gives
\begin{equation}
-\prod_{p \in A} Z_p \prod_{q \in B} Z_q g_k.
\end{equation}
Note that the first qubit is $Z_1$ because the edge $(1,m)$ is toggled twice, and hence is still on. Thus, $H_1 Z_1$ acts non-trivially, and we get
\begin{equation}
g_m' = -(X_1 Z_1) \prod_{p \in A} Z_p \prod_{q \in B} Z_q g_m
\end{equation}
and so (since $X_1$ anticommutes with $Z_1$):
\begin{equation}
    g_1' g_m' = X_1 \prod_{p \in A'} Z_p g_m = g_1 g_m.
\end{equation}

This proves \cl{3}. Now we will prove the $g_i'$ stabilize $\ket{\psi}$. First, rewrite $\ket\psi = (I + g_1') \ket{G}$.
Thus, $g_1'$ stabilizes $\ket{\psi}$ because $g_1' (I + g_1') = I + g_1'$.
Also, for $m \notin A$, we have $g_m'$ stabilizes $\ket{\psi}$ because $g_m'$ commutes with $g_1'$ and
\begin{equation}
    \fontsize{10pt}{10pt}\selectfont g_m' \left( I + g_1' \right) \ket{G} = \left( I + g_1' \right) g_m' \ket{G}
    = \left( I + g_1' \right) \ket{G}.
\end{equation}
Finally, for $m \in A$,
\begin{equation}
    \fontsize{9pt}{9pt}\selectfont g_m' \left( I + g_1' \right) \ket{G} = \left( I + g_1' \right) g_1' g_m' \ket{G}
    = \left( I + g_1' \right) \ket{G}.
\end{equation}

This completes $k = 0$.

\subsection{The case $k = 1$}
Suppose $k=1$. We first note that
\begin{multline}
    i\prod_{q \in B} Z_q \ket{G} = -Y_1 X_1 \prod_{q \in B'} Z_q \ket{G} \\
    = -Y_1 \prod_{q \in B'} Z_q \prod_{p \in A'} Z_p \ket{G}\\
    = -Y_1 \prod_{q \in B} Z_q \prod_{p \in A} Z_p \ket{G}.
\end{multline}
We have the following claims.
\begin{enumerate}
    \item $\displaystyle g_1' = -Y_1 \prod_{q \in B} Z_q \prod_{p \in A} Z_p$.
    
    \item If $m \notin A$, then $g_m'$ stabilizes $\ket{G}$.
    
    \item If $m \in A'$, then $g_1' g_m'$ stabilizes $\ket{G}$.
\end{enumerate}

We show $\displaystyle\left(I - Y_1 \prod_{q \in B} Z_q \prod_{p \in A} Z_p \right) \ket{G}$ is stabilized by $g_m'$, in a similar way to the case $k = 0$.

Before doing the computations, we observe that $S$ commutes with $Z$, so the only non-trivial action of $S$ is converting an $X$ to a $Y$. This we represent by $Y_m X_m$ when appropriate.

The proof of \cl{1} follows the same way as in the $k = 0$ case.
\begin{multline}
    g_1' = (Y_1 X_1) \cdot \prod_{p \in A'} Z_p \\
    \cdot \prod_{p \in A'} Z_p \prod_{q \in B'} Z_q \cdot (-1) \cdot X_1 \prod_{p \in A'} Z_p \\
        = -Y_1 \prod_{q \in B'} Z_q \prod_{p \in A'} Z_p 
        = -Y_1 \prod_{q \in B} Z_q \prod_{p \in A} Z_p
\end{multline}
The second line is the action of $\zeta$ on $g_1$, and the first is the action of $\sigma$ and $H_1 Z_1$. The only $X$ that an $S$ acts on is $X_1$, which we write as $Y_1 X_1$. $H_1 Z_1$ commutes with $Y_1$ so $H_1$ acts trivially.

\cl{2} follows the same way as when $k = 0$.

\cl{3} has two cases: $m \in B$ and $m \notin B$.

For $m \notin B$, the action of $\zeta$ gives $\prod_{q \in B} Z_q g_m$, while the action of $\sigma$ gives
\begin{equation}
    (Y_m X_m) \prod_{p \in A\backslash\{ m \}} Z_p \prod_{q \in B} Z_q g_m.
\end{equation}
Note that the operation on the first qubit is $Z_1$ because the edge $(1,m)$ is toggled twice, so it is on. Thus, $H_1 Z_1$ acts non-trivially, and
\begin{equation}
    g_m' = (X_1 Z_1 Y_m X_m) \prod_{p \in A\backslash\{ m \}} Z_p \prod_{q \in B} Z_q g_m.
\end{equation}
As a result,
\begin{multline}
    g_1' g_m' = -Y_1 \prod_{q \in B} Z_q \prod_{p \in A} Z_p \\ (X_1 Z_1 Y_m X_m) \prod_{p \in A-\{ m \}} Z_p \prod_{q \in B} Z_q g_m = g_m.
\end{multline}

For $m \in B$, the action of $\zeta$ on $g_m$ gives $\prod_{p \in A} Z_p \prod_{q \in B} Z_q g_m$, while the action of $\sigma$ gives
\begin{multline}
    -(Y_m X_m) \prod_{p \in A\backslash\{ m \}} Z_p\cdot \prod_{p \in A} Z_p \prod_{q \in B} Z_q g_m \\
        = i \prod_{q \in B} Z_q g_m.
\end{multline}
Note that the operation on the first qubit is $I_1$ because the edge $(1,m)$ is toggled thrice, but was on originally (since $m \in A$) and hence is now off. Thus, $H_1 Z_1$ acts trivially, $g_1' = i \prod_{q \in B} Z_q g_m$, and
\begin{equation}
    \fontsize{11pt}{11pt}\selectfont g_1' g_m' = -Y_1 \prod_{q \in B} Z_q \prod_{p \in A} Z_p 
         \cdot i \prod_{q \in B} Z_q g_m = g_1 g_m.
\end{equation}

\subsection{Same global phase}
We now prove that the two sides of Equation \ref{initassume} have the same global phase.
Let $c = \frac{1}{\sqrt{2^n}}$. We apply $\bra{0} = \bra{00\dots0}$ to both sides. Also define $\bra{1} = \bra{10\dots0}$. On the LHS, the $\prod Z$ acts on the $\bra{0}$ trivially, since $\ket{0}$ is a $+1$ eigenstate of $Z_p$ for all $p$.
Furthermore, since
$\displaystyle\ket{G} = \prod_{(a,b)\in E} CZ_{a,b} \ket{+}$
where $E$ is the edge set of $G$, and since $CZ_{a,b}$ stabilizes $\bra{0}$ and $\bra{1}$, we have
\begin{equation}
    \bra{0}\prod_{(a,b)\in E} CZ_{a,b}\ket{+} = \braket{0}{+} = c,
\end{equation}
\begin{equation}
    \bra{1}\prod_{(a,b)\in E} CZ_{a,b}\ket{+} = \braket{1}{+} = c.
\end{equation}
Thus, $\braket{0}{G} = \braket{1}{G} = c$, so
\begin{equation}
    \braopket{0}{\frac{1}{\sqrt{2}} \left( I + i^k \prod Z \right)}{G} = \frac{1 + i^k}{\sqrt{2}} \cdot c.
\end{equation}
For the RHS, we note that $\ket{0}$ is a $+1$ eigenstate of $Z$, $CZ$, and $CS$. We then have
\begin{equation}
    \braopket{0}{Z_1 \prod CS_{x,y}^k \prod CZ_{x,y}}{G} = c.
\end{equation}
\begin{equation}\label{blah}
    \braopket{1}{Z_1 \prod CS_{x,y}^k \prod CZ_{x,y}}{G} = i^k \cdot c.
\end{equation}
This is since $\ket{10\dots0}$ is a $-1$, $i$, and $+1$ eigenstate of $Z_1$, $S_1$, everything else, respectively. Note the two $Z_1$'s in the above product; one written as $CZ_{1,1}$, which verifies Equation \ref{blah}.
Thus,
\begin{equation}
    \fontsize{11pt}{11pt}\selectfont\braopket{0}{H_1 Z_1 \prod CS_{x,y}^k \prod CZ_{x,y}}{G} = \frac{1 + i^k}{\sqrt{2}} \cdot c.
\end{equation}
In summary, we proved that applying $\bra{0}$ to both sides gives the same nonzero result (since $k \neq 2$), which means the global phases are equal.
This proves the correctness of our algorithm.

\section{Runtime}
Consider the action of a $\calC_3$ gate $C$ on stabilizer state $U \ket{G}$. The steps of our algorithm are:
\begin{itemize}
    \item Express $C$ as a sum of two products of projectors $D + E$ and $\calC_2$ gates, as in Table \ref{decomp}.
    
    \item Form the conjugate $D' = U^{-1} DU$.
    
    \item Convert all $X$'s and $Y$'s to $Z$'s.
    
    \item Apply the merging algorithm in Proposition \ref{merge}.
\end{itemize}
The main contributors to the runtime are the last two steps. If $n$ is the number of qubits and $d$ is the average degree of $\ket G$, we claim that our average runtime is $O(nd)$.
The first two steps above take $O(n)$ time since $|\langle H,S\rangle|=24$.
Converting an $X$ or $Y$ on a qubit to $Z$'s, which must be done for each of $n$ qubits, requires $d$ applications of $Z$ on average, hence has $O(nd)$ runtime.
Lastly, applying the gates in the algorithm takes $O(|A|\cdot|A|+|A|\cdot|B|)$.
However, on average we have $|A|=d$ and $|B|=n$, so we get $O(d^2+dn)$.
Since $d<n$, we get runtime $O(nd)$.

\section{Conclusions}
We showed how to simulate $\calC_3$ circuits on stabilizer states. We use a decomposition of $\calC_3$ operators into a sum of projectors and a merging algorithm that combines two states. We showed robustness of our algorithm with circuit identities, and we also derived a decomposition of the magic state $TT\ket{++}$ in two stabilizer states, in accordance with the known results. As further research, we can look at how our algorithm performs when computing higher-qubit magic states. We can also look to derive more cases where merging two states is possible besides the ones we found. Finally, we can look at the implications of our $\calC_3$ circuit simulation algorithm for universal quantum computation.

\section{Acknowledgements}

The authors are grateful to Professor Isaac Chuang for suggesting the research question and fruitful discussions as well as to Professor Peter Shor for helpful comments and encouragement.

\end{document}